# Influence of cognitive, geographical, and collaborative proximity on knowledge production of Canadian nanotechnology

Elva Luz Crespo Neira[1,*], Ashkan Ebadi[1,3], Catherine Beaudry[2], and Andrea Schiffauerova[1]

[1] Concordia Institute for Information Systems Engineering (CIISE), Concordia University, 1515 Ste-Catherine Street West, Montréal, Québec H3G 2W1, Canada

[2] Department of Mathematical and Industrial Engineering, École Polytechnique de Montréal, 2900 Edouard Montpetit Blvd, Montréal, Québec H3T 1J4, Canada

[3] National Research Council Canada, Montréal, Québec H3T 2B2, Canada

**Abstract**

Incorporating existing knowledge is vital for innovating, discovering, and generating new ideas. Knowledge production through research and invention is the key to scientific and technological development. As an emerging technology, nanotechnology has already proved its great potential for the global economy, attracting considerable federal investments. Canada is reported as one of the major players in producing nanotechnology research. In this paper, we focused on the main drivers of knowledge production and diffusion by analyzing Canadian nanotechnology researchers. We hypothesized that knowledge production in Canadian nanotechnology is influenced by three key proximity factors, namely cognitive, geographical, and collaborative. Using statistical analysis, social network analysis, and machine learning techniques we comprehensively assessed the influence of the proximity factors on academic knowledge production. Our results not only prove a significant impact of the three key proximity factors but also their predictive potential.

## 1. Introduction

Nanotechnology is an emerging scientific field that targets solid particles in the size range of 1-100 nm (Yegul, Yavuz, & Guild, 2008). Scientific publications in nanotechnology have substantially increased since the decade of the 1990s, with nanoscience programs flourishing throughout the world (Delemarle *et al.*, 2009). Additionally, the involvement of academic research in nanotechnology is preeminent. With universities as an important source for knowledge generation and even innovative activities, nanotechnology appears to rely on sciences to a higher degree than other technology fields (Wang & Guan, 2011).

Canada is one of the top players in nanotechnology in terms of the number of scientific publications (Yegul *et al.*, 2008). Moreover, Canadian nanotechnology inventors show an increasing tendency to collaborate more closely with other researchers (Beaudry & Schiffauerova, 2011) which may foster the scientific output (Ebadi & Schiffauerova, 2016a) and knowledge diffusion among the scientific community. Sharing of knowledge in academic research is essential for innovation to take place (Moodysson & Jonsson, 2007). The mutual exchange of knowledge and shared learning indicates that knowledge is intrinsically a socially constructed process (Berger & Luckmann, 1991). Several studies analyzed the role of collaboration networks and their structural properties on the spread of knowledge and technological development (Eslami, Ebadi, & Schiffauerova, 2013; Schilling & Phelps, 2007) and found a positive relation.

---

[*] Corresponding author, elvitaluz@gmail.com, +593 98 508-0830

Proximity is defined as the existing interactions between actors (Gilly & Torre, 2000). There exist several proximity dimensions in the literature (Boschma, 2005; Gilly & Torre, 2000; Moodysson & Jonsson, 2007; Zeller, 2004). Although there are some discrepancies towards the number of proximity dimensions, the factors to contemplate are common. Rather, they are sometimes renamed, or just either separated into several or merged in one dimension. For explaining and understanding how connections are encouraged into the formation of networks, it is useful to design a framework constituted by multiple proximity dimensions. In this research, we focused on three dimensions of proximity, *i.e.* cognitive, geographical (Boschma, 2005), and collaborative. Cognitive proximity is defined as the shared knowledge base and expertise of different entities, also labeled as actors (Boschma, 2005). Interdisciplinary research collaboration can be related to the cognitive proximity, when visualizing the interactions of scientists from complementary fields has proven helpful to obtain an overview of interdisciplinarity in academic research outputs (Boschma & Frenken, 2010).

Geographical proximity is the most common dimension (Knoben & Oerlemans, 2006) and is defined as the spatial or physical distance between actors, either measuring it with an absolute metric of length, *i.e.* actual distance units, or a relative metric, *e.g.* travel time (Boschma & Frenken, 2010). Although spatial concentration could bring significant benefits, having too much geographical proximity between actors has its own risks for knowledge production, and innovation. One potential risk is that actors within a region could become a too "inward-looking" community with weakened learning ability, and lose their capacity to come up with new ideas or respond to new developments (Gittelman, 2007). Several studies argue that spatial proximity no longer matters as digital means are substituting more and more the need for co-location. In the current global economic world, distance is not as indispensable as much as it used to be, back when the information and communication technologies (ICTs), as well as transportation means, were not as developed (Boschma, 2005; Bouba-Olga & Ferru, 2012; Sonn & Storper, 2008). Although on one hand, it is unquestionable that scientists can communicate well over long distances, the impact of the internet can be limited due to the tacit nature of knowledge itself, and the social nature intrinsic in the innovation and learning processes (Feldman, 2002).

Scientific collaboration in academic research is a complex social phenomenon (Glänzel & Schubert, 2004) that scholars have been trying to measure ever since the 1960s (De Solla Price, 1965). Collaborative proximity represents the distances between scholars and their positions within academic collaboration networks. A scientific network may not be the best tool for quantifying a purely social component (Newman, 2001). The reason is scientific networks do not directly measure actual contacts between researchers, though their structure would reflect the society that built them. We coined the term *collaborative proximity*, by virtue of adjusting to the proper terminology in the proximity frameworks. If we assume that original ideas arising from academic publications or technological innovation are the product of collaborations between actors (Eslami *et al.*, 2013), collaboration is thence key for knowledge production. Moreover, there is an increasing interest in using natural mechanisms of cognition and information filtering for scholarly purposes through social connections, such as collaboration links. Co-authorship networks are a proxy measure adequate to quantify collaboration among groups of researchers. Two scientists who have worked together at least once are more likely to later keep in touch for meaningful information exchange, and thus, it would presumably result in repeated collaboration between researchers (Agrawal *et al.*, 2006; Beaudry & Schiffauerova, 2011).

Even though there is yet little understanding of how proximity affects innovation over time (Boschma, 2005), it is considered as an influencing factor for knowledge flows in science. Notably, former research on nanotechnology suggests that proximity plays a significant role in the understanding of collaboration and development in this area (Cunningham & Werker, 2012). A certain degree of proximity is required to make the actors well-connected in scientific collaboration networks. One may expect increasing proximity to result in more interactions between actors, leading them, in turn, to learn and innovate more. However, proximity acts as a double-sided blade. According to the *proximity paradox* concept of Boschma and Frenken (2010), whereas too little proximity may prevent interactive learning and innovation from happening, too much proximity can be also harmful. Hence, an optimal level of proximity between actors needs to be reached and not surpassed, to avoid negative impacts on their academic and innovative performance, due to the lack of openness and flexibility (Boschma, 2005; Broekel & Meder, 2008).

Having known that proximity may impact the knowledge flow, the question is how to track something as intangible as knowledge? Trajtenberg *et al.* (1997) affirm that even though knowledge flows are invisible, leaving no trail by which they may be measured and their patterns discerned, they do leave a paper footprint in the form of citations. Citations have been extensively applied for studying knowledge diffusion across a variety of dimensions, and are valid measures for tracing the knowledge flows (Alcacer & Gittelman, 2006). Additionally, several studies analyzed collaborative proximity and observed significant correlations between various network structure measures and researchers' performance (Abbasi, Altmann, & Hossain, 2011; Ebadi & Schiffauerova, 2016b; Onel, Zeid, & Kamarthi, 2011; Wallace, Larivière, & Gingras, 2012). One should note that citing behavior would mostly suggest the endorsement, authority conferral, provenance tracking, and scholarly trust (Ding, 2011). However, citation counts do not have the potential of yielding insights into the motives behind an author's citing behavior (Bornmann & Daniel, 2008). Since citations are sometimes made for social reasons (Shadish *et al.*, 1995), citing behavior may be a simple indicator of more complex behaviors or social relationships (Lievrouw, 1989). Thus, among the factors that influence citing behavior, social and cognitive reasons are considerably involved.

Although several studies used citation and collaboration network analyses to evaluate knowledge production, many of them have either only employed bibliometric approaches, or specifically focused on patent data. It becomes evident that the literature predominantly deals with proximity concerns involved in an innovative citation, or even its collaboration with scholarly communities but without really exploring their impact on the citation; studies on scientific citation are, as a matter of fact, rather scarce. Further, the influence of geography has only been examined on academic citation when innovation is implicated. To the best of our knowledge, there is no study dealing with geographical, cognitive, and collaborative dimensions combined in a comprehensive analysis. Additionally, although there are some studies applying collaboration network analysis to Canadian scientific publications, they are either outside the realm of nanotechnology, focusing on other scientific fields (*e.g.* Sarigöl *et al.*, 2014); plus, sometimes the research was more geared towards business economics (*e.g.* Schummer, 2004), or performed at a macro level (e.g. Delemarle *et al.*, 2009). Additionally, scientific works dealing with social network analysis on nanoscience knowledge production are aggregated analyses that consider academic networks only from a high-level perspective (*e.g.* Onel *et al.*, 2011), or have a very limited scope (*e.g.* Rafols & Meyer, 2007). This is the first study that comprehensively analyzed the impact of proximity on knowledge production of Canadian nanotechnology and inspected knowledge-generating networks at the micro-level of the academic community, exploring the attributes of its individual members.

In this study, we address three main questions in the field of Canadian nanotechnology: 1) Is cognitive proximity between two scholars a strong influencing factor as to increase the probability of an existing citation link among them?, 2) Does geographical proximity between two researchers have an impact sufficient to result on an increased probability for the establishment of a citation link among them?, and 3) Does collaborative proximity, as measured by the location of two authors within a co-authorship network, have an effect on the probability of a citation link between them? Based on these research questions, we form three hypotheses:

- **H$_0$ 1:** Citation probability does not increase when a Canadian author is cognitively proximate to another author.
- **H$_0$ 2:** Citation probability does not increase when a Canadian author is geographically proximate to another author.
- **H$_0$ 3: a)** The probability of a Canadian author citing another does not increase due to the referenced author's position within the collaboration network. **b)** The probability of a Canadian author citing another does not increase if they are closely connected to each other within the collaboration network.

The remainder of the paper proceeds as follows: *Data and Methodology* section describes the data and techniques in more detail; the *Results* section presents the research topics and their trends; the paper concludes in the *Conclusion* section and some future directions and limitation of the research are discussed in the *Limitations and Future Work*.

## 2. Data and Methodology

In this section, the data and methodology are discussed in detail. The main part of the analytics pipeline was coded in PHP, Javascript, and Python. We also used some statistical modeling, and network analysis software, which are specified later in the respective section. The final dataset was stored in a MySQL database.

### 2.1. Nanotechnology publications

The scope of this research covers the knowledge production of Canadian nanotechnology researchers. Scientific publications have largely been accepted as the easiest and most relevant measure of scientific knowledge production (Callon *et al.*, 1986; Delemarle *et al.*, 2009). However, as an emerging and highly interdisciplinary field, there are some concerns about a proper data set of nanotechnology publications (Delemarle *et al.*, 2009; Schummer, 2004). Among the main challenges are the lack of a specific tag for nanotechnology in many conventional databases, such as the web of science (WoS), and the fact that words including the term "*nano*" do not exclusively refer to the nanoscience publications (*e.g.* "*nanokelvin*"). We followed the approach introduced in Moazami *et al.* (2015) that addressed the mentioned challenges and extracted nanotechnology publications from Elsevier's Scopus. The data collection method incorporated specialized keywords related only to nanotechnology to select relevant academic papers, while also filtering out those with misleading terms such as "*nanosecond*". This data set was cross-referenced with the WoS database to obtain further paper details, such as the scientific field of articles. The database also includes specific details about publications and authors, such as the location of the authors and the article's publication year. We used citation links as the distinction of the type of relationship, similar to the work by Cantner *et al.* (2013) on patent applications. Therefore, we defined two types of authors: 1) Citing Canadian (CC) authors who published at least one Canadian paper within the examined time interval, and 2) Authors of the cited papers (REF), who may or may not be cited by CC authors and can be either Canadian or non-Canadian. We assumed that a "*Canadian paper*" is an article with at least one Canadian author among the authors. Author affiliation was used as a proxy for the geographical location to identify Canadian researchers. We extracted 3,981 articles published by Canadian nanotechnology researchers (CC authors) during 2010 and 2011.

We decided to evaluate the citing behavior of these CC authors against researchers from three previous years respectively, by selecting cited papers published within a five-year time window (that is, between 2007 and 2010). In other words, we paired up CC authors from 2010 against REF authors from 2007 to 2009, and CC authors from 2011 against REF authors from 2008 to 2010. We also dismissed year pairs such as <CC-2011, REF-2011>, and <CC-2010, REF-2010>, so that no pair was from the same year. Thus, our REF authors data was comprised of the authors from 395,051 cited papers, making a total of 467,794 cited scholars.

Having collected the data, the next step was pairing CC and REF authors. We realized that during the period of 2007 to 2010, only 26,987 REF authors had at least one citation connection to a Canadian author from 2011 and 2010, while the majority (*i.e.* 440,807 authors) did not have this citation link. Therefore, the data was highly imbalanced as the number of cited authors with a positive citation link was only 5.77%. We used the synthetic minority over-sampling technique (SMOTE) to overcome the imbalanced data problem (Chawla *et al.*, 2002). This algorithm is a powerful sampling technique that oversamples rare events by creating additional synthetic observations of that event, while at the same time under-sampling the population majority that does not contain the desired effect. It has been previously used in various fields and applications including link prediction in social networks (Munasinghe & Ichise, 2011; Wang *et al.*, 2007). After oversampling, we obtained a set of 27,014 non-linked authors vs 26,987 linked authors (50.02% vs 49.98%).

Using the balanced data, we then created all the combinations between the CC and REF authors. Since we had no specified path between the non-linked REF authors and Canadian authors, to pair these up we used an algorithm to randomize these combinations, whereas, for the positively linked scholars, we followed the respective citation records. Although authors favor their own papers for citing, which has been deemed as a beneficial factor in citation counts by others (Bethard & Jurafsky, 2010), we ignored all the self-citing links as they were not interesting for our research objectives. Besides, self-citations are naturally more geographically localized (Jaffe & Trajtenberg, 1999), thus including them could favorably bias the effect of the spatial dimension. We also filtered out authors with no affiliation data available so we could be able to measure the geographical proximity. The final dataset contained 80,091 author pairs, from which 41.8% (33,477 combinations) were positively linked, that is, with an actual citation connection between them, and 58.2% (46,614 combinations) had a negative, or non-citing connection (or "control" citation links). The 80,091 author pairs consisted of 2,824 Canadian papers published between 2010 (59,621 pairs) and 2011 (20,470 pairs), authored by 3,747 distinct Canadian scholars (CC authors), and 34,877 reference papers published between 2007 and 2010, authored by 47,380 distinct REF authors.

### 2.2. Measuring proximity

*2.2.1. Measuring cognitive proximity*

The common practice for measuring cognitive proximity is using field or subject categorization in a scientific database (*e.g.* Boufaden & Plunket, 2007; Jaffe & Trajtenberg, 1999; Rafols & Meyer, 2010). Similarly, our dataset contained information about the journal that the article was published in, the article's domain, field, and subfield, which helped us to identify its scientific field.

Identifying the cognitive category would be also helpful in assessing the degree of interdisciplinarity through creating a categorical scale of "*closeness*" between the cognitive fields and subfields of each author in the pairing. We defined three levels for the cognitive category as follows:

- Category 1: Same subfield and same field as well
- Category 2: Only same field (different subfields)
- Category 3: Different fields

We assumed that the fields and subfields of an article apply to all the authors, even though the field and subfield information in the database pertain to the paper, rather than specifically to the author. Based on the IDs of both the article and the reference (cited) paper, we fetched the cognitive field and subfield information corresponding to the publication. Next, we coded a program to compare the fields and subfields between the two authors and calculate the cognitive distance according to our predefined scale. Finally, we split the variable into three dummy variables for each category.

*2.2.2. Measuring geographical proximity*

We followed Cunningham and Werker (2012) in measuring the geographical proximity by decomposing it into: 1) physical proximity (*i.e.* to Euclidean or spatial distance), and 2) functional proximity (*i.e.* the traveling time existing between two authors). Euclidean distance refers to the straight-line distance between two points in the Euclidean space, in this case, locations of the two scholars, denoted in kilometers. It should be noted that this measurement is in "*as the crow flies*" or "*in a beeline*" kilometers, meaning that it is the shortest distance between two geographical coordinates on a map, disregarding terrain considerations. Since our data set did not contain postal codes, we could not use geocoding software to easily convert them into coordinates. Therefore, we first extracted the affiliation address of each author. Next, a program was coded to pass this address information to the web services offered by Google Maps, the Geocoding API[†], that converted addresses into geographic coordinates. Finally, we obtained the Euclidean distance by using the Haversine formula[‡], which calculates the distance in kilometers between two points on a sphere from their longitudes and latitudes through considering the radius and curvature of the Earth.

For the functional proximity, we estimated the traveling time between scholars. We considered two different methods of transportation, plane and vehicle, thus dividing time into two sub-metrics: 1) Flying time, and 2) Driving time. These two metrics were later combined in a single variable: traveling time. The driving time was only calculated when the Euclidean distance between the two authors was less or equal to 500 kilometers (Km). In the literature, some authors studying the influence of spatial distance on collaboration in other countries used the threshold of 100 miles (~160 Km) as the maximum distance indicator for preferring travel by car over a flight (Bouba-Olga & Ferru, 2012; Laursen *et al.*, 2011). However, in Canada cities are farther from each other, and researchers may drive for distances greater than 100 miles, for instance, the distance from Montreal to Quebec City is ~250 Km (160 miles), or from Montreal to Toronto is ~500 Km (310 miles). Therefore, we set the limit at 500 Km as the maximum distance for which scholars would choose to drive. For all remaining distances greater than 500 Km, we assumed that the author would take an aeroplane to reach the destination. We used the already calculated Euclidean distance to measure the flying time, and as for speed, we considered the Boeing 747 aircraft as the airplane model with a cruising speed of 885 Km/h[§]. Although we acknowledge that for local flights the aircraft model might be different, it still serves as a useful baseline for reference. Next, we coded a program to calculate the flying time in "hh:mm:ss" time format. We further adjusted the flying time factor by allocating reasonable extra time to account for the time it takes to get to the airport, average waiting times before a flight, and to get from the arrival airport to the destination (Torres *et al.*, 2005). Finally, we combined both traveling sub-metrics, obtaining a consolidated metric for traveling time.

---

[†] Google Inc. Google maps web service APIs, 2015. URL: https://developers.google.com/maps/web-services/.
[‡] GeoDataSource.com. Distance calculation in php. https://www.geodatasource.com/developers/php, 2015. Accessed: 2015-03-12.

Having calculated the traveling time, the last step was converting the variable to a categorical variable. Similar studies used a ranking approach for categorizing the geographical distance, using geopolitical boundaries (*e.g.* Cunningham & Werker, 2012). We followed the approach and categorized the traveling time according to the difference or likeness in the geopolitical location between the two scientists in the dyad. We defined five levels for the location categorical variable:

- Category 1: Same city or town
- Category 2: Same state or province
- Category 3: Same country
- Category 4: Same continent
- Category 5: Different continent

Finally, we coded a program to compare the geopolitical fields between the two authors and assign the location category according to the mentioned scale. As we did with the cognitive distance, we also transformed the location category into five "*dummy*" variables.

*2.2.3. Measuring collaborative proximity*

We created an undirected co-authorship network for the period of 2007 to 2011, in which authors are represented as nodes and they are linked if they have a joint paper, and used social network analysis (SNA) to estimate the collaborative proximity between researchers. The extracted network was highly connected with 94.87% of the nodes being part of the largest component. We used Pajek version 4.06 (Batagelj & Mrvar, 1998) and NetworkX version 1.10 (Hagberg, Swart, & S Chult, 2008), which is a network analysis package in Python programming language, to inspect the network structure and obtain the centrality measures. All the parameters measuring individual characteristics of a node pertain to the reference node (REF), *i.e.* the cited author. The only exception is the shortest path, which instead refers to pairs of scholars. In the following section, we briefly introduce the calculated social network measures.

*2.2.3.1. Shortest path*

The shortest path represents how many authors are between two given authors tied up by co-authorship links with their peers (Newman, 2004). For the case when the two specific scholars have collaborated in a publication, they would be direct neighbors, and the shortest path would be zero. On the other hand, there might be no path between the nodes at all, in which case we set a large value, *i.e.* 9999, representing infinity (Newman, 2004).

---

[§] Microsoft Encarta Encyclopedia. 747 jumbo jet, 2000. URL:
http://hypertextbook.com/facts/2002/JobyJosekutty.shtml. CDROM. Accessed: 2015-03-16.

*2.2.3.2. Degree centrality*

The degree centrality of a node is defined as the number of nodes that are connected to that node. This measure is the sum of all the node's directly connected neighbors and is then normalized between zero and one.

*2.2.3.3. Betweenness centrality*

Betweenness centrality measures how often a node is located on the shortest path between other nodes in the network. Betweenness centrality is normalized as the proportion of all geodesics that include that particular node, thus it can also be expressed as a percentage. A node's betweenness centrality would express its average capacity to control the flow of information in the network (Ebadi & Schiffauerova, 2015; Uddin *et al.*, 2012). The idea of this centrality measure is that if a node with a high level of betweenness was removed, the network would fall apart into coherent clusters (Leydesdorff & Zhou, 2007). Hence, nodes acting as proxies to join different clusters would have a high betweenness value.

*2.2.3.4. Closeness centrality*

Closeness centrality is the average of the shortest path distances between a specific node and all the other nodes in a network. Hence, this value would only be available for connected nodes. It expands on the concept of degree centrality by emphasizing how close a node is to all other nodes in the network (Uddin *et al.*, 2012). As a result, the more central a node is, the lower its total distance from all other nodes is going to be.

*2.2.3.5. Eigenvector centrality*

Eigenvector centrality estimates the centrality for a node based on the centrality of its neighbors basing on the eigenvector of the largest eigenvalue of an adjacency matrix. This centrality metric relates to the influence of a node in the network by assigning relative scores to every node. The idea behind this concept is that links to nodes representing authors with high scores contribute more to the score of that particular authors than links to low-scoring authors. Unlike degree centrality, which considers every node equally, eigenvector centrality weighs nodes according to their centralities. Since this means considering not only direct connections but indirect links as well, eigenvector would be a centrality measure considering the entire network pattern (Bonacich, 2007).

*2.2.3.6. Clustering coefficient*

The clustering coefficient measures the degree of the clustering tendency of a given node in a network, by making use of the links to all the other nodes in the network. In a co-authorship network, the clustering coefficient of a node represents the willingness of this node's collaborators to collaborate with each other, indicating the probability that two of its collaborators wrote a paper together (Barabási *et al.*, 2001).

## 2.3. Methods and variables

We study the impact of proximity on academic knowledge production through the performance measure of an existing citation link in a dyad, which is a pair of authors. Scientific citation probability can either be formulated as a statistical regression problem, or alternatively, as a classification problem to be solved with supervised learning techniques (Dong *et al.*, 2015; Jawed *et al.*, 2015; Yan *et al.*, 2012). The citation relationship is extracted from the pairings, with this link expressed as a binary value and taken as the dependent variable. Given the Boolean nature of our dependent variable, we adopted two methodologies to conduct the analysis:

1. Logistic regression modeling

2. Machine learning classification modeling

Each method took the citing behavior as the dependent variable, that is if a citation link has been established or not. A set of independent variables based on the metrics representing collaborative proximity and general network attributes, as well as those representing cognitive and geographical proximity were considered. The list of independent variables are as follows:

1. Cognitive proximity
    a. Cognitive category
2. Geographical proximity
    a. Euclidean distance
    b. Traveling time
    c. Location category
3. Collaborative proximity
    a. Degree centrality
    b. Betweenness centrality
    c. Closeness centrality
    d. Eigenvector centrality
    e. Clustering coefficient
    f. Shortest path

We also considered two control variables that might have an important effect on citation probability, *i.e.* published works and co-authored works. The published work variable refers to the number of articles published by the target or cited author during the specified time period. Since we consider that a researcher that has published a high number of papers could be deemed prestigious in the field, we expect the author to be highly cited by peers regardless of their proximity measurements. The co-authored works variable represents a simple count of the published works that have been co-authored by each pair of authors. It is likely that if two researchers have co-published scientific articles, they cite each other in their subsequent works. Having defined the variables, we performed intensive preprocessing including but not limited to dealing with duplicate pairs and data quality checks, making the data ready for analysis.

*2.3.1. Logistic Regression analysis*

We used Stata[**], a data analysis and statistical software, to test the validity of our hypotheses. We suspected some of our independent variables to be interrelated, such as travel time and Euclidean distance, or that high collinearity may exist between some centrality metrics (Valente *et al.*, 2008). Therefore, we used Pearson product-moment correlation coefficient (PC) to check correlations among the variables. Conventionally, variables are uncorrelated if $PC < 0.1$, weakly correlated if $0.1 < PC < 0.3$, moderately correlated if $0.3 < PC < 0.5$ and strongly correlated if $0.5 < PC < 1.00$ (Schiffauerova & Beaudry, 2011). Figure 1 shows the correlation plot. All the correlations were statistically significant at the level of 0.05. The cognitive category and shortest path were found to have a negative moderate correlation to the citing response, whereas the other proximity metrics displayed a weak correlation. In terms of the centrality metrics, we found the weak linear correlation unsurprising (Sarigöl *et al.*, 2014). For the control variable, coauthored works were found to have a weak correlation to the shortest path, whereas published works were strongly correlated to the target author's degree centrality, as well as to betweenness and eigenvector centrality.

---

[**] StataCorp. Stata statistical software: Release 14.1, 2015.

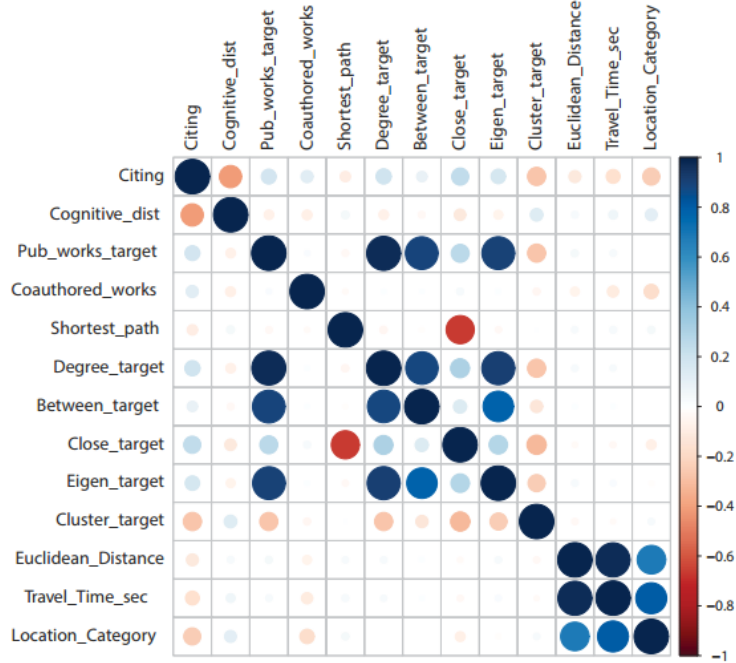

**Figure 1.** Correlation among dependent and independent variables

We predictably found a strong correlation between degree and betweenness centrality, and to eigenvector and closeness centrality. The clustering coefficient was in general weakly correlated to the rest of the SNA metrics, except for a negative moderate correlation to the shortest path. Closeness centrality and shortest path had a negative moderate correlation while being moderately correlated to degree centrality and clustering coefficient, and a positive moderate correlation to location category. Moreover, as anticipated, there was a high correlation between Euclidean distance, traveling time, and location category. Based on the analysis, we omitted the published works control variable as well as closeness centrality. For the remaining centralities, *i.e.* degree, betweenness, and eigenvector, we decided to run the regression model using each one of them alternately at a time. In the same way, we assessed the effects of the geographical metrics, *i.e.* Euclidean distance, traveling time, and location category, by including them in the model by turns. Given that the impact of spatial distance may be better comprehended in combination with other factors (Morgan, 2004), we also included an interaction variable of Euclidean distance and shortest path in the model. Other potential interactions were not found significant.

We also performed goodness-of-fit (GOF) tests, namely, the Hosmer-Lemeshow and $R^2$ tests. The GOF tests were not satisfactory, however, it has been previously suggested that these tests might not perform well every time as opposed to tests targeting linear models (Hosmer *et al.*, 1997; Le Cessie & Van Houwelingen, 1991). We suspected that this issue might be due to our large sample size. Therefore, we decided to run an experiment with a smaller random sample. We discovered that for regressions using less than 1,000 observations (we tested with a 1% random sample of about 700 observations), the fitting of our model was found to be significant. Thus, we proceeded to run the experiment 10 times, each time with a different 1% random sample. All the proximity variables remained significant throughout each run, except for the coauthored works variable. In most runs, the effect of this control variable was omitted due to "*separation or quasi-separation*", that refers to a scenario in which the dependent variable does not change by different levels of the independent variables (Long & Freese, 2006). In this case, having co-published at least one paper would ensure the existence of a citation link, which is the expected behavior for direct co-authors.

Nevertheless, we observed that pairs with co-authored works have a very low incidence in our data, only 1,798 pairs (2.26%).

*2.3.2. Machine learning classification*

As part of our work, we implemented cross-validated machine learning classification to be able to evaluate the performance of the model, prevent over-fitting, and validate the results on a disjoint unseen set of data. For this purpose, we first split the data into 60% and 40%, representing the training data and the validation set respectively. Next, we trained the model on the training data by applying a 10-fold cross-validation design. In a 10-fold cross-validation, the data (here we refer to the training data) is first divided into ten subsets, taking nine subsets as training and one set as the test set, and then in ten separate iterations, ten models are trained on the training data and tested on the test set. Once all ten models are built and evaluated, the best model is selected. We also performed hyperparameter tuning to calibrate the parameters of the machine learning classifier. We ran the model in two different settings, one including a feature selection module called the reduced model, and the other with all the independent variables.

We tested various machine learning classifier models in our data, finally obtaining the best results with Random Forests (RF). Originating from *decision trees*, RF is an ensemble technique that grows forests of trees and lets them vote for the most prominent class (Breiman, 2001). Notably, RF is considered as a high-performing classifier, robust against over-fitting, that has shown comparable or even better prediction performance than other learning methods (Breiman, 2001; Xu, 2013). It has been extensively used in classification problems in various domains and applications (e.g. Dong *et al.*, 2015; Lichtenwalter *et al.*, 2010; Sarigöl *et al.*, 2014).

## 3. Results

In this section, we present the results of the regression and machine learning models that were applied to predict the establishment of a citation link.

### 3.1. Logistic regression model

Using various combinations of independent variables, accounting for the scientific domain, geography, and network centrality, as well as the spatial-social interaction, we ran a logistic regression model to check for statistical dependency between citing behavior. Even after having controlled for the co-publishing effect, all variables in the model were found to be statistically significant at the confidence level of 95% (Table 1). One may suspect that the high significance was obtained due to the large number of observations, yet the high z-score values would indicate otherwise (Singh, 2005), thus confirming the effectiveness of the predictors. As seen in Table 1, the model achieved ~76% accuracy and precision, with the area under the curve (AUC) of higher than 81%. We considered the prediction cutoff point at 0.5.

**Table 1.** The logistic regression model results

|  | Variable | Coefficient | z-score | R Std. Err. |
|---|---|---|---|---|
| **Cognitive** | cog1 - same subfield | 2.32 | 22.79 | 0.10 |
|  | cog2 - same field | 1.31 | 13.41 | 0.10 |
| **Geographical** | Euclidean distance | -1.85E-04 | -9.15 | 1.94E-05 |
|  | Traveling time | -4.95E-08 | -13.54 | 3.66E-09 |
|  | loc1 - same city/town | 2.90 | 11.91 | 0.24 |
|  | loc2 - same state | 0.95 | 6.22 | 0.15 |
|  | loc3 - same country | 0.87 | 7.97 | 0.11 |
|  | loc4 - same continent | 0.38 | 4.78 | 0.08 |
| **Collaborative** | Degree centrality | 0.01 | 12.15 | 0.00 |
|  | Betweenness centrality | 1.38E+04 | 12.59 | 1.10E+03 |
|  | Eigenvector centrality | 2.69E+02 | 10.59 | 2.54E+01 |
|  | Clustering coefficient | -0.94 | -13.25 | 0.07 |
|  | Shortest path | -0.61 | -6.83 | 0.09 |
| **Interaction variable** | Euclidean-Shortest path | 3.22E-05 | 4.94 | 4.19E-06 |
|  |  |  |  |  |
| **Performance (%)** | **Accuracy** | 76.21 |  |  |
|  | **Precision** | 76.30 |  |  |
|  | AUC | 81.72 |  |  |

Note: 1) Closeness centrality was not included in the model, as discussed in the data and methodology section.
2) All variables are significant at the confidence level of 95%.

We ranked the predictors according to the z-score, as a measure of their association with the dependent variable (Thompson, 2009). This metric does not depend on the unit of measure of the parameters, as standardized coefficients do in the logistic regression model. Since some factors had negative effects, *i.e.* traveling time, Euclidean distance, shortest path, and clustering coefficient, we used the absolute value of the z-score. As seen in Table 1 and considering all the independent variables, cognitive proximity was the most influential factor in the regression, followed by collaboration, and the geographical aspect. For the cognitive proximity, we included the scientific domain effect in the regression by using the two dummy variables (cog1 and cog2). The high z-scores of the cognitive variables (22.79 and 13.41) are an indication of their high contribution for establishing citation links. In fact, the magnitude of having the same subfield was the greatest factor among the explanatory variables.

For the geographical proximity, we observed that being in the same city/town (loc1) is decisive when it comes to establishing a link between CC and REF authors, closely followed by being in the same country (loc3) and the same state (loc2), and to a lesser degree, being within the same continent (loc4). Additionally, both Euclidean distance and traveling time presented a negative effect on the outcome, meaning that they increase citation probability by decreasing in value. Regarding the collaborative proxies, surprisingly a negative effect was observed for the clustering coefficient. Recall that high clustering coefficient values refer to nodes at well-connected "neighborhoods"; hence, its negative influence could possibly mean that REF authors located in the middle of collaborator cliques are not being as highly cited. We considered the eigenvector centrality, as a metric that better reflects a scholar's central position regarding the entire network structure. Thus, it can be said that CC authors would have a citing preference towards REF nodes either with a high number of collaborators (degree) or strategically located as proxy nodes between different clusters (betweenness) but not necessarily embedded in the midst of a highly-cliquish collaborator neighborhood.

Furthermore, the shortest path reflected a negative contribution, which implies that there is indeed a certain preference for citing REF nodes that are closer to the CC authors by acquaintanceship; however, its effect is not as high. Presumably, it is because we are facing a small-world network, with almost 95% connected authors (the largest component), overall clustered at 75%, and with a mean of almost 4 hops between dyads. Therefore, the collaborator-chain factor, although important in the model, is not decisive to establish a citing link. Finally, the spatial-social interaction effect created by combining Euclidean distance and the shortest path was found significant. Thus, we confirm what we suspected about the shortest path, in that it sometimes has the capacity to substitute for spatial proximity, with the latter serving as its proxy. From the results, it seems the structure of the co-authorship network, as measured by various collaborative metrics, is meaningful to the academic citation, the same as the cognitive proximity.

### 3.2. Machine learning classifier

The performance evaluation of the random forests classifier on the disjoint validation set validated the high predictive power of the trained machine learning model. The model achieved a promising performance of 86.91% accuracy, 85.95% precision, and AUC of 93.19%. We also checked for the relative importance of the predictors based on the "mean decrease in accuracy" metric. As seen in Figure 2, the cognitive proxies and the shortest path are listed as the top three predictors. Interestingly, almost similar importance is observed for the collaborative factors, measures by the network centrality measures. However, eigenvector and degree centrality contributed slightly better to the model than betweenness, while traveling time marginally outperformed Euclidean distance, as also seen in the regression. The location category has the lowest ranking in the RF model.

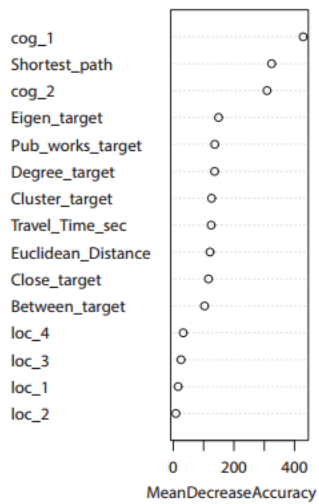

**Figure 2.** Ranked list of the independent variables based on their importance in the random forests model, using the "mean decrease in accuracy" and all the independent variables.

As explained in the data and methodology section, we also ran the model with a feature selection module (the reduced model). In the reduced model with six selected variables, we observed that their importance ranking does not seem to vary much when compared to the complete model (Figure 3). The cognitive measures and the shortest path were still the top predictors, proceeding with the collaborative measures.

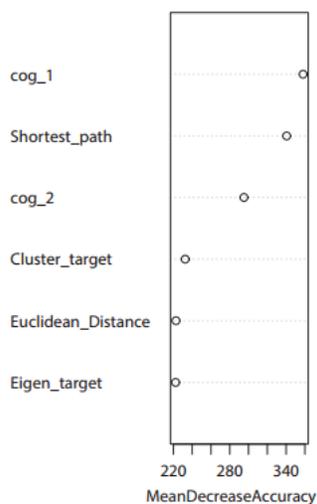

**Figure 3.** Ranked list of the independent variables in the reduced model based on their importance in the random forests model, using the "mean decrease in accuracy".

### 3.3. Hypotheses testing

Based on the results, in this section we check the hypotheses stated earlier concerning the impact of proximity aspects on the probability of scholarly citation.

*$H_0 1$: Citation probability does not increase when a Canadian author is cognitively proximate to another author.*

Cognitive proximity was found statistically significant in both the logistic regression and the random forests models. Moreover, it was proven to be the main contributor to the citation effect by both statistical and machine learning approaches; CC scholars will decisively cite REF authors publishing in the same subfield as themselves, and will also have a preference towards those from their own field, as opposed to academics publishing outside of it. Consequently, we reject $H_0 1$ and instead, we assert that cognitive proximity is, in fact, an influential factor for citing

*$H_0 2$: Citation probability does not increase when a Canadian author is geographically proximate to another author.*

We had enough evidence to reject the hypothesis to reject $H_0 2$ as well, on the grounds of the significance of all the metrics representing spatial proximity. Not only, were Euclidean distance and traveling time important to increase citing likelihood, both as alternate stand-alone factors as well as a combined effect with dyad collaborative distance, but they also proved to have predicting capability towards it. Importantly, authors co-located within the same country (loc3), same state (loc2), and city (loc1) are very likely to be linked by citation.

*$H_0 3a$: The probability of a Canadian author citing another does not increase due to the referenced author's position within the collaboration network.*

Next, we examined the hypotheses concerning collaborative proximity and an academic's position within the co-authorship network. We attempted to verify whether the citation probability increased the more central a cited author was found to be. Essentially, the clustering coefficient along with the various centrality metrics turned out to be significant, which gave us sufficient evidence to reject $H_0 3a$, having also displayed prediction power for the establishment of a citation link. This finding would imply that Canadian scientists are more likely to cite a REF author that has a good position within the collaboration network, in terms of: 1) having co-published papers with a high number of people (degree), 2) whether he is more central in the entire network (eigenvector), 3) if the scholar acts as a proxy for clusters within the network (betweenness), and if the author is situated outside of highly-cliquish neighborhoods (clustering coefficient).

*$H_0$ 3b: The probability of a Canadian author citing another does not increase if they are closely connected to each other within the collaboration network.*

Based on the findings on the shortest path variable, we rejected $H_0$ 3b hypothesis in preference of the alternate statement: "The closer two scholars are located with respect to each other within the co-authorship network (through the chain of co-authoring peers), the higher the probability is for a positive citing link between them." Furthermore, the shortest path would interact with Euclidean distance to add up to the explanatory power of the model, while also using the latter as a proxy to establish such acquaintanceship connections, and thus substitute for spatial closeness. Having collaborative closeness would thus greatly serve to anticipate a citation link between authors, with the shortest path being among the best-performer variables in the citing prediction.

## 4. Conclusion and Discussion

We evaluated the importance of three key factors for the creation of knowledge in nanotechnology in Canada. To the best of our knowledge no study has considered all cognitive, geographical, and collaborative proximities to comprehensively explore the production of academic information. We generated a feature framework based on measurements from these factors, based on statistical and classification approaches to assess their influence on effective citations.

Our results suggest that coming from close cognitive bases makes authors more prone to be linked by citation. This is in line with Ding (2011) that analyzed academic citations in the field of information retrieval and indicated productive authors prefer peers sharing their same research field to both cite and coauthor with. Additionally, Hu and Jaffe (2003) similarly found that technological proximity is important for knowledge flows, as evidenced by increasing patent counts. Likewise, Jaffe and Trajtenberg (1999) had claimed that technologically proximate inventors are preferred for citing, with Cunningham and Werker (2012) affirming the same behavior for patenting firms. It can be said that cognitive interests are important for citing behavior, be it academic or innovative. Moreover, we showed that despite nanotechnology commonly being depicted as a highly multidisciplinary field, cognitive similarity still matters. In fact, Schummer (2004) had previously stated that nanotechnology displays a merely average multidisciplinarity level in its scientific citation patterns, not differing much from other scientific fields.

However, previous research indicates that interdisciplinarity is primordial in the development of nanotechnology (Malsch, 1997) for contributing to the academic society in terms of integrating knowledge from assorted domains. Also, since articles with an interdisciplinary background are allegedly more successful due to having increased value for the scholarly community (Katz & Martin, 1997), interdisciplinarity is an important factor for knowledge production. Besides, high levels of it in research are closely related to innovation (Rafols & Meyer, 2010; Weingart, 2000), and, on a practical note, it has been brought up that programs funding nanoscale research usually take interdisciplinary approaches (Schummer, 2004). Overall, it seems Canadian nano-scientists would have to balance the need of being cognitively proximate and try to aim for the "idealistic nano-visions" (Rafols & Meyer, 2007) of producing more interdisciplinary knowledge, in light of its weighty benefits for knowledge production and innovation. This could be achieved by seeking information from complementary cognitive domains, thus avoiding the potential issues coming from cognitive lock-in that Boschma (2005) warns about.

In similar studies analyzing the effect of collaborative proximity, researchers typically focused on the effects of one or two centrality measures, such as degree centrality (*e.g.* G. Wang & Guan, 2011) and/or shortest path. We examined the impact of collaborative proximity on citations by constructing a comprehensive framework which accounted for various collaborative metrics. While our results are in line with previous findings about the fact that having a high degree centrality is favorable for citation purposes (either academic or innovative) (*e.g.* Eslami *et al.*, 2013; Wallace *et al.*, 2012), there are wide discrepancies regarding the other centrality measures. Liu *et al.* (2014) argued that while the degree is the best-performing centrality metric, betweenness centrality only evidenced a mild effect on the scholarly citation. Similarly, further analyses suggested that solely by combining academic and innovation networks of collaboration, *i.e.* when scientific research translates into practical applications, the power of betweenness centrality to control knowledge flows revealed (Eslami *et al.*, 2013). Furthermore, in the study by Abbasi *et al.* (2011), both betweenness and closeness centralities were not statistically significant. As for eigenvector centrality, whereas Liu *et al.* (2014) deemed it as non-significant, Abbasi *et al.* (2011) claimed that it had a negative effect on the scholarly citation. Contrarily, our results suggest otherwise, except for closeness centrality, which was dropped from our statistical modeling. In this respect, although not evidenced by regression, the closeness centrality was found almost equally important as all other centrality metrics in the RF model. Overall, it can be said that a nanotechnology research paper is more likely to be cited if the authors are better located in the collaboration network. Although the impact of the shortest path was not the highest among the factors in the regression model, our results support the general agreement that citing likelihood decreases as the shortest path increases (Singh, 2005; Sorenson, Rivkin, & Fleming, 2006). This finding contrasts with Wallace *et al.* (2012), who claimed that apart from direct self-citations, they did not find a strong preference to cite authors close in the co-authorship network. The fact that the random forests model assigned a higher weight to the SNA metrics, along with their statistical significance in the regression model, suggests that collaboration is decisive when it comes to picking academic references. Particularly, the effects of the centrality metrics along with the clustering coefficient show the influence of how central authors are more likely for citing.

It has been argued (Bouba-Olga and Ferru, 2012) that geographical proximity only matters when innovation, a knowledge-intensive economic activity, is involved. In our work, each of the geographical metrics was found significant by the regression machine learning models, manifesting a valid statistical dependence between spatial considerations and citation probability. Thus, our work constitutes, to our knowledge, the first one that addresses scholarly citation without any involvement from innovation literature or sources. Our findings comply with conclusions from patent analyses, which assessed different scales of location category and their influence to establish a citing link (Breschi & Lissoni, 2003; Singh, 2005), in that geography is not a sufficient condition for knowledge diffusion, usually requiring interaction from other aspects, like collaborative or cognitive proximity, as well. Other studies manifest the same for collaboration intensity as represented by patent citation count and frequency, although only the country level was considered (Hu & Jaffe, 2003; Jaffe & Trajtenberg, 1999). To support our findings about geographical proximity, we further evaluated the citation network, as represented by the precise locations of positively linked scientists (Figure 4). In the figure, the greater the node's in-degree (meaning an author being highly cited by others), the bigger the node is, and the darker its color.

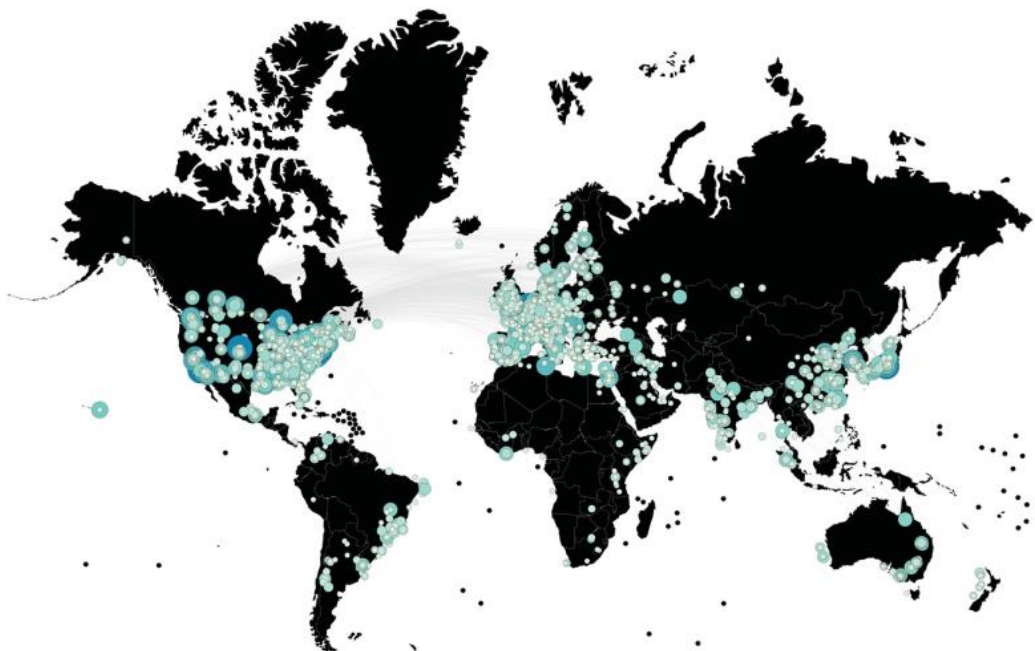

**Figure 4.** Geo-layout of authors with positive citing links

A high density of incoming citations in North America, all over Europe, and in Asia near China, is observed while citations are sparser in other places. Thus, from the perspective of Canadian scientists, although spatial co-location may be important, it would not make much difference in where the cited author is located, as compared to the richness of the source in terms of the scientific field. Lastly, despite Canada's place among leading countries in nanotechnology, it would still be surpassed by conglomerate areas of knowledge with a major concentration in the field, located in the US, Europe, and Asia.

Last but not the least, many studies have been conducted with positive citation links, usually, in terms of citation counts, whereas we used both negative and positive linkages, and we employed a prediction approach as well. Thus, we remark on the use of machine learning models to study the behavior of the proximity factors and discovering how strong a predictor they can be for citation probability. Research combining citation analysis and machine learning usually comes from other knowledge bases, such as physics (Lichtenwalter *et al.*, 2010), and computer science (Bethard & Jurafsky, 2010; Dong *et al.*, 2015). We did not find any paper focusing on the academic papers on nanotechnology nor addressing any combined proximity concerns.

We expect our findings to be encouraging for nanotechnology researchers, motivating them to pursue trans-disciplinary research, that is, making use of the best research ideas and methodologies from other fields. In addition, we anticipate that observable effects from the scholarly collaboration will drive institutions to promote the establishment of social connections in this academic domain. Policymakers shall also gain result-based evidence to support fine-tuning of science and technology policies that encourage multidisciplinary research teams. We anticipate that this study will provide supportive arguments for the creation of tools that will facilitate the expansion of research fields within Canada.

## 5. Limitations and Future Work

We were exposed to some limitations. We considered a 5-year window for the co-authorship networks, implying the very real possibility that two authors may have co-authored a paper outside this time range. As a consequence, in our network they appear as disconnected, affecting the SNA metrics and shortest path, likely leading to isolated clusters. Surely, extending the collaboration network by more years would derive in more relationships between the authors and a denser network. However, many authors were found to be part of the largest component of the network, thus this limitation did not turn out to be highly detrimental to our work. In addition, there is a chance that successful scholars (in terms of published articles) may have stopped publishing in year one of our focus period, due to passing away or being retired, for example. In this case, the high degree that would account for receiving citations from Canadian researchers would be missing, whereas if we considered previous years, such author would be more central. Therefore, our analysis could be improved by incorporating larger time-windows.

Another limitation was the lack of address information in the database for some scientists. This resulted in the weeding out of a portion of our observations, because otherwise not every case would have had the geographic attributes required for our analysis. We suspect that the referenced scientific databases do not include affiliation details for all the possible co-authors in an article. Hence, although the proportion of discarded pairs was low, we advise caution to this potential bias in the interpretation of our results related to spatial distance.

We considered papers with at least one Canadian author as *Canadian papers*, without accounting for the fact that although one of the authors may have a Canadian address, the research may have been majorly conducted elsewhere. This restriction would be better examined through co-authoring pairs rather than citation dyads. Finally, we focused on the nanotechnology domain as an emerging field. As a consequence, any collateral effects due to different citation patterns from different disciplines were dismissed, although at the expense of generalization. However, in this limitation, we also find an opportunity for future research.